\documentclass[
preprint,
 amsmath,amssymb,
 aps,
floatfix,
]{revtex4-1}

\usepackage{graphicx}
\usepackage{dcolumn}
\usepackage{bm}

\begin{document}

\title{Impact of chemical short-range order on radiation damage in Fe-Ni-Cr alloys}

\author{Hamdy Arkoub}
\author{Miaomiao Jin}
\email{Corresponding author: M. Jin (mmjin@psu.edu)}
\affiliation{Department of Nuclear Engineering, The Pennsylvania State University, 218 Hallowell Building, University Park, 16802, PA, USA}
\begin{abstract}
Chemical short-range order (CSRO), a form of nanoscale special atom arrangement, has been found to significantly alter material properties such as dislocation motion and defect dynamics in various alloys. Here, we use Fe-Ni-Cr alloys to demonstrate how CSRO affects defect properties and radiation behavior, based on extensive molecular dynamics simulations. Statistically significant results are obtained regarding radiation-induced defect propensity, defect clustering, and chemical mixing as a function of dose for three CSRO levels. The perfect random solution as an energetically unfavorable state (negative stacking fault energy) shows the strongest tendency to enable diffusion, while a high CSRO degree scenario generally reduces the effective defect diffusivity due to trapping effects, leading to distinct defect dynamics. Notably, in the high-CSRO scenario, interstitial clusters are Cr-rich and interstitial loops preferentially reside in/near the Cr-rich CSRO domains. It is also identified that CSRO is dynamically evolving in a decreasing or increasing manner upon continuous irradiation, reaching a steady-state value. These new understandings suggest the importance of incorporating the effect of CSRO in investigating radiation-driven microstructural evolution.

\begin{description}
\item[Keywords]
Fe-Ni-Cr alloys, chemical short-range order, radiation damage, defects,  molecular dynamics

\end{description}
\end{abstract} 
\pacs{Valid PACS appear here}
\maketitle


Fe-Ni-Cr-based austenitic steels are commonly used in nuclear applications as structural materials. Radiation-induced elemental redistribution in the alloys due to radiation-enhanced diffusion and ion mixing (ballistic and thermal spike mixing) \cite{was2016fundamentals} may induce chemical short-range order (CSRO) as observed in concentrated alloys \cite{chen2021direct}. Although compositional variation has long been recognized to cause phase instability \cite{was2016fundamentals}, the potential formation of CSRO has not been emphasized adequately, particularly considering its impact on defect dynamics \cite{cao2021does}, which constitutes the basis of microstructural evolution. 

CSRO has been commonly reported in multiple principal element concentrated alloys such as medium or high entropy alloys (MEA/HEA). Due to the perturbations of electronic structure and phonon spectrum of the constituent species, the atomic configuration exhibits some atomic correlations that can be described as CSRO \cite{gomez2014decoupling,chen2021direct,zhang2020short}. CSRO has been used to explain the increased mechanical characteristics in various alloys \cite{chen2021direct,zhang2020short,su2022radiation,li2019strengthening,jian2020effects}. The degree of CSRO may be tailored to achieve an optimized combination of strength, ductility, and toughness via control of composition and thermoprocessing parameters \cite{zhang2020short,jian2020effects,ding2018tunable}. For example, Jian et al. \cite{jian2020effects} identified a large degree of CSRO increases the resistance to nucleation and migration of Shockley partials in CrCoNi alloys, increasing the yield strength. Moreover, CSRO can significantly modify the stacking fault energy (SFE) due to its dependence on the local atomic environment \cite{zhao2019local,smith2016atomic}. The ground state SFE of CrCoNi-based random solutions was calculated to be near-zero or negative \cite{zhang2017dislocation,zhang2017origin,zhao2017stacking,pei2021hidden}, contrary to experimental measurements \cite{laplanche2017reasons,okamoto2016size}. It was later demonstrated that the SFE of CrCoNi MEA can vary broadly by tuning the degree of CSRO \cite{ding2018tunable}, and increasing CSRO leads to a higher SFE \cite{li2019strengthening,ding2018tunable}. As SFE is fundamentally related to dislocation properties, the impact of CSRO on SFE in Fe-Ni-Cr systems needs to be evaluated.

Not only does CSRO pose significance in modulating dislocation migration, it essentially affects the kinetic behavior of general defects due to the modified potential energy landscape (PEL). Cao tracked the point defect migration in molecular dynamics simulations and identified that the defect diffusivity is reduced in thermally annealed CrCoNi with CSRO \cite{cao2021does} compared to random solid solution. It was speculated that the CSRO should also impede dislocation loop and grain boundary migration \cite{cao2021does}, e.g., the 1D glide of defect clusters was observed in Ni but not in the CrCoNi alloy (note: no direct confirmation of CSRO in this work, but CSRO is expected based on the fabrication method) \cite{wang2019defect}. Therefore, CSRO is likely to have a substantial role in affecting defect properties and dynamics in Fe-Ni-Cr alloys under irradiation. Thermodynamically, CSRO is enhanced by atom diffusion, which is a thermally slow process due to a low concentration of defects. By contrast, radiation-enhanced diffusion can promote an efficient increase of CSRO \cite{gomez2014decoupling,zhang2017local,su2022radiation} over the timescale of radiation experiments. However, how the CSRO correlates with radiation dose and the existence of steady-state CSRO remain to be elaborated. 

To reveal the impact of CSRO on the radiation damage in Fe-Ni-Cr alloys, we consider Fe$_{0.2}$Ni$_{0.5}$Cr$_{0.3}$ with an experimentally stable face-centered cubic (FCC) lattice \cite{bonny2013interatomic}, where a high concentration of Cr ($>$10-11\%) is known to induce Cr clustering \cite{gomez2014decoupling,allen2008radiation}. Hence, the Cowley short-range order parameter $\alpha$ \cite{cowley1950approximate} is used to indicate the degree of Cr clustering in a quasi-binary alloy:  
$\alpha_i = 1 - n_i/(c N_i)$, where $n_i$ is the number of the non-Cr atoms among the $N_i$ atoms
in the $i$th shell, and $c$ is the nominal concentration of the non-Cr atoms (=0.7). Here, we use the first nearest neighbors (1NNs) for CSRO calculations, given the pertinent role in quantifying CSRO experimentally \cite{zhang2017local,chen2021direct}. We use molecular dynamics/statics to compute defect energetics and simulate the radiation process. Fe-Ni-Cr atomic interactions are described by a commonly used EAM potential by Bonny et al. \cite{bonny2013interatomic}, which reproduces well the stacking fault energy and vacancy formation energy and provides acceptable values for the elastic constants for stainless steels.

\begin{figure*}[htbp]
  \centering
  \includegraphics[width=0.6\linewidth]{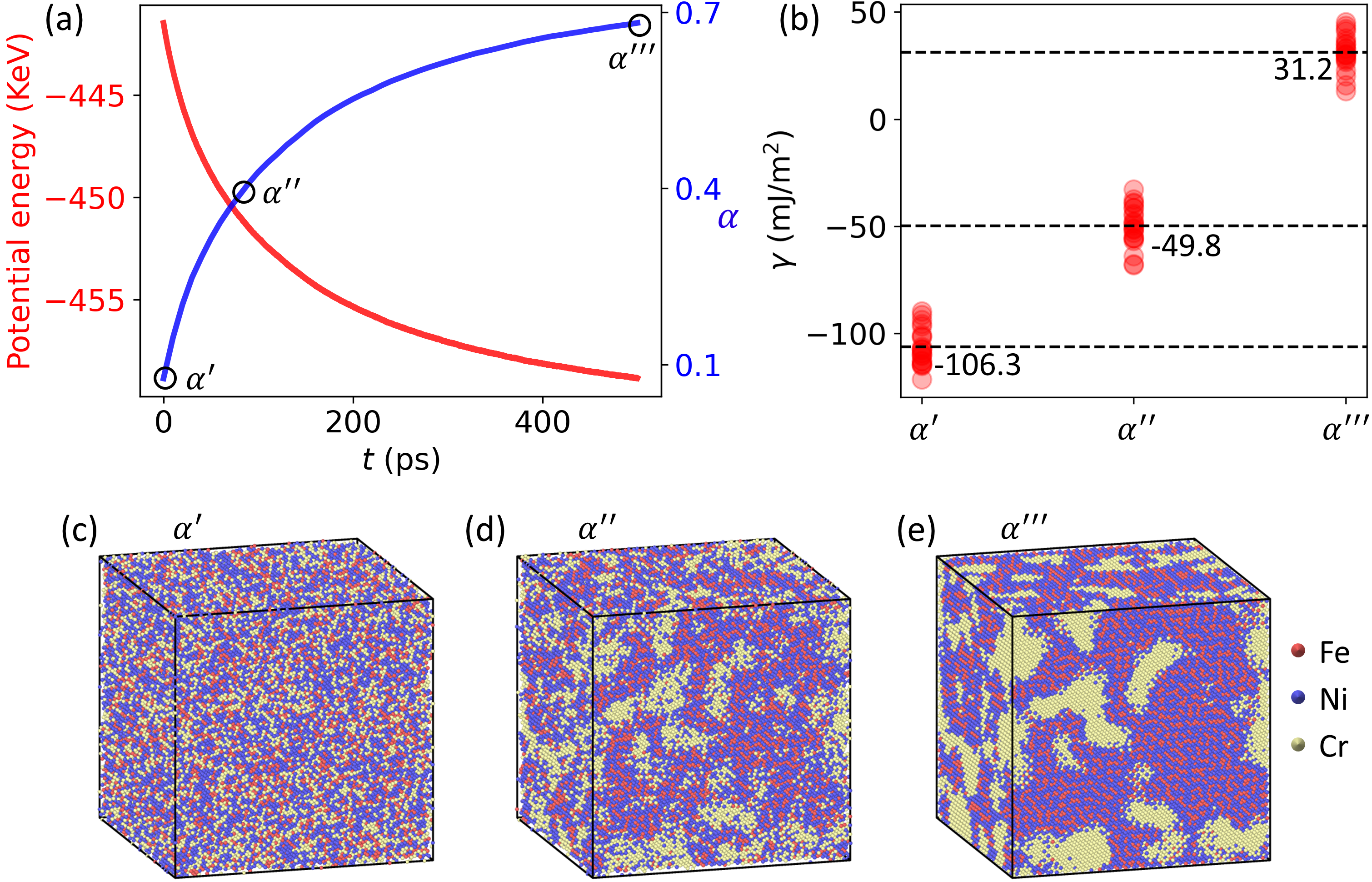}
  \caption{(a) Potential energy and the CSRO parameter with increasing MCMD timesteps. (b) SFE for three selected CSRO degrees, denoted as $\alpha'$, $\alpha''$, and $\alpha'''$ in (a). (c-d) the corresponding atom configurations to $\alpha'$, $\alpha''$, and $\alpha'''$, respectively.}
  \label{fig:mc-ordering}
\end{figure*}
With a random solution configuration (108,000 atoms, around $11\textrm{nm}\times11\textrm{nm}\times11\textrm{nm}$), metropolis Monte Carlo molecular dynamics (MCMD) gradually evolves the system to a thermodynamically favorable configuration, under zero pressure at 300 K with a time step 0.001 ps. FIG. \ref{fig:mc-ordering}a plots the change of total potential energy and calculated CSRO $\alpha$ versus MCMD steps. The increase in $\alpha$ is commensurate with the decrease in potential energy. To investigate the CSRO impact, we select three configurations corresponding to low ($\alpha'=0.076$), medium ($\alpha''=0.39$), and high ($\alpha'''=0.68$) CSRO degrees, shown in FIG. \ref{fig:mc-ordering}c-e, respectively. The apparent Cr clustering is consistent with previous experimental and computational studies \cite{lavrentiev2007monte,gomez2014decoupling,wang2021predicting,li2013atomic}.

Given the importance of SFE in explaining radiation-induced defects, dislocations, and phase stability, the SFE is calculated for the three CSRO degrees. To compute SFE ($\gamma$), a [111] oriented cell is created at the specified CSRO. An intrinsic stacking fault is created by shifting the top half $a/6 \langle 112\rangle$ with the bottom half fixed, where $a$ is the lattice parameter. The SFE is calculated as the energy cost to create the SFE divided by the stacking fault area. For statistical quality, we perform such a shift at 45 different layers for each CSRO (more details in supplementary materials (SM)). The values of SFE are summarized in FIG. \ref{fig:mc-ordering}b. Due to the local variation of composition, slight spread is identified in each scenario, while drastic change occurs as CSRO increases: the average $\gamma$ increases from -106.3 mJ/m$^2$ to 31.2 mJ/m$^2$ as CSRO increases from  $\alpha'$ to  $\alpha'''$.  Here, the positive value associated with a high CSRO degree is within the range of experimental measured SFEs for Fe-Ni-Cr alloys 13-57 mJ/m$^2$ \cite{miodownik1978calculation}. It is hence possible to establish a correlation between CSRO and SFE, such that SFE is tunable by manipulating the CSRO level, consistent with the understanding of MEA/HEA systems \cite{zhang2020short,li2019strengthening,ding2018tunable}.
\begin{figure*}[htbp]
  \centering
  \includegraphics[width=0.6\linewidth]{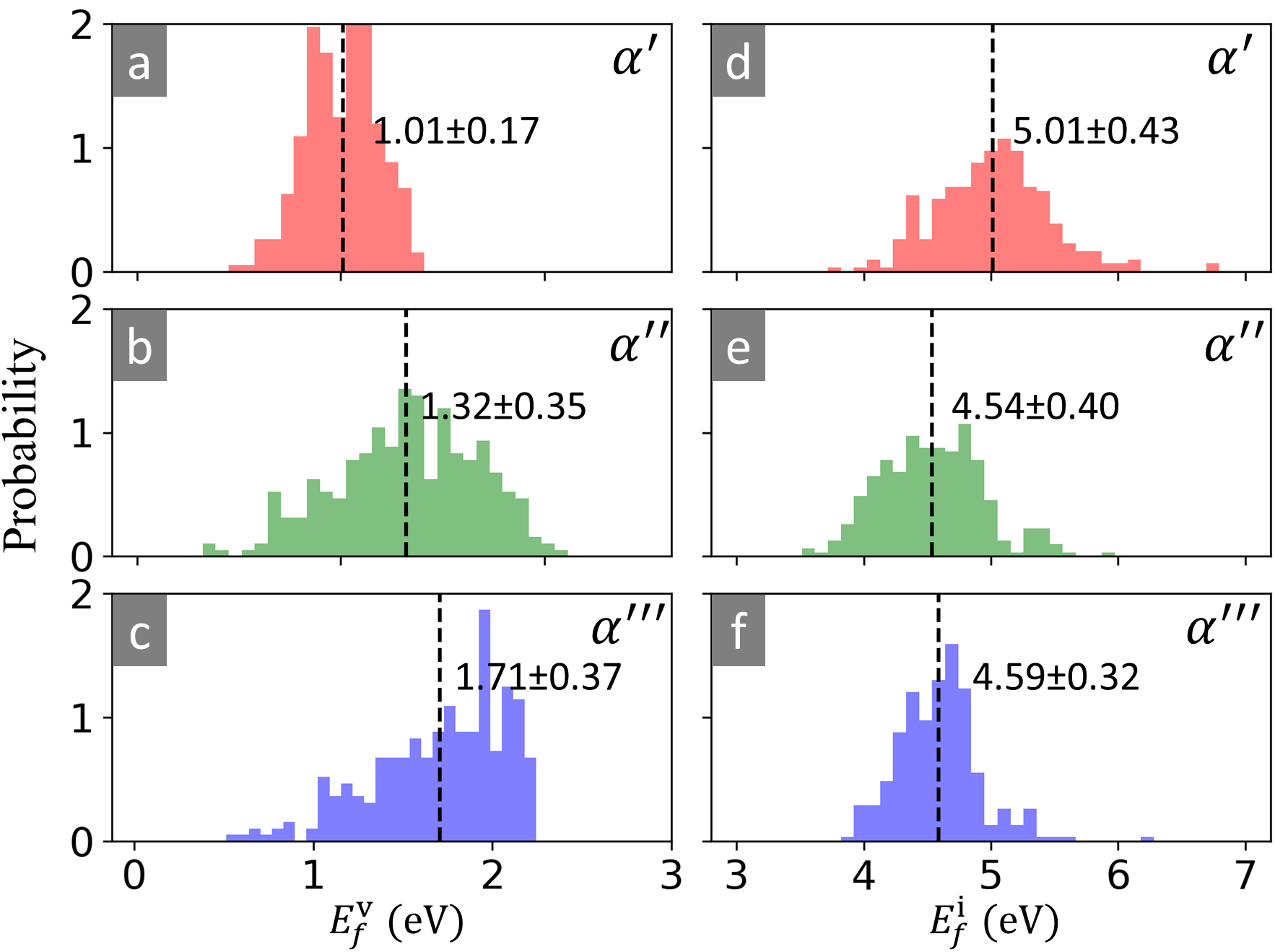}
  \caption{(a-c) Vacancy formation energy ($E_f^\mathrm{v}$) and (d-f) interstitial formation energy ($E_f^\mathrm{i}$), corresponding to CSRO $\alpha'$, $\alpha''$, and $\alpha'''$, respectively.}
  \label{fig:formation}
\end{figure*}

To probe the radiation-induced defects, we first calculate the point defect properties for each CSRO to reveal changes in point defect formation and migration energetics. Due to the local lattice distortion from the alloying elements, we consider 100 instances of randomly located point defects to generate the formation energy spectra. Cr vacancy is created by removing one Cr atom. FIG. \ref{fig:formation}a-c indicate that the formation energy shifts to the higher energy range as CSRO increases, which is attributed to Cr in the Cr-rich CSRO domain with reduced lattice distortion. For Cr interstitial, the structure is more complex given the local atomic environment. Hence, Cr-Cr dumbbells along [111], [100], and [110] directions are individually created, and after energy minimization, most initial configurations relax to non-low-index directions (see stereo plots in SM). From \ref{fig:formation}d-f, the formation energy of Cr-Cr dumbbell is reduced in high CSRO scenario ($\alpha'''$) due to the CSRO domains.

\begin{figure*}[htbp]
  \centering
  \includegraphics[width=0.75\linewidth]{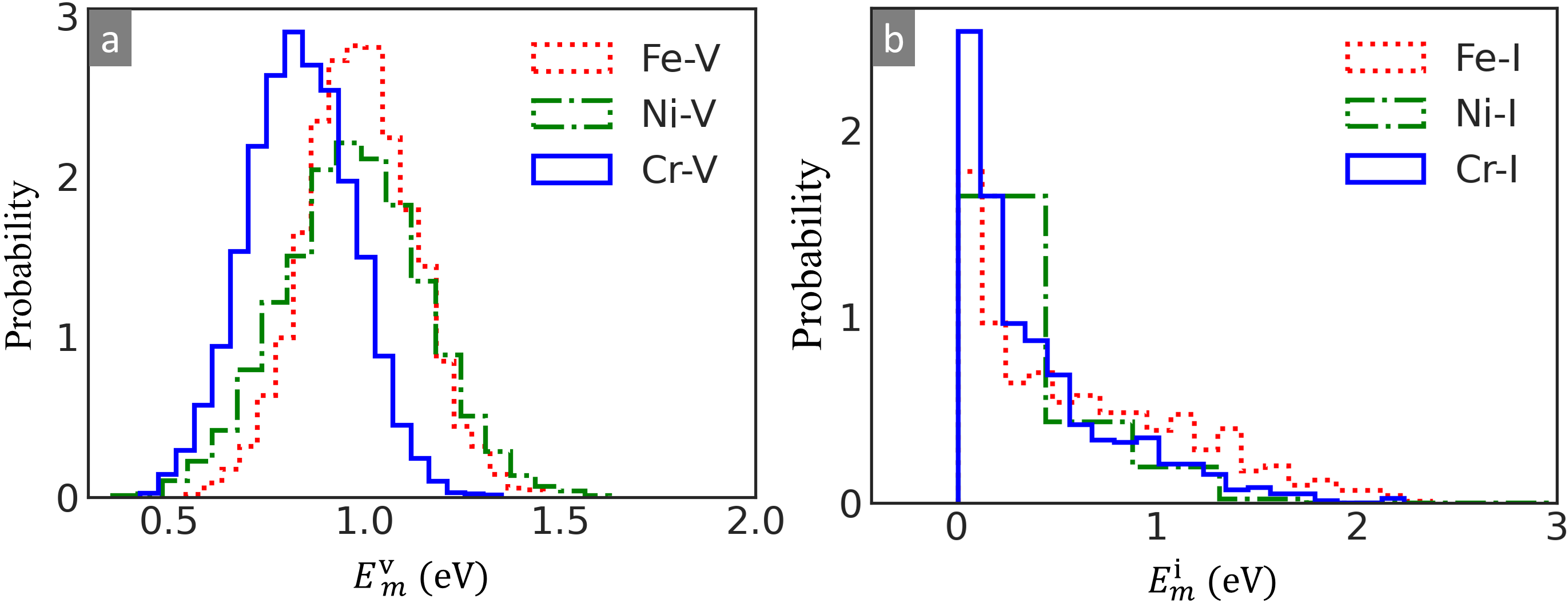}
  \caption{Migration barriers of vacancy (a, $E_m^\mathrm{v}$) and interstitial (b, $E_m^\mathrm{i}$), in random solution.}
  \label{fig:migration}
\end{figure*}

To see how CSRO changes the migration energetics, we compare the migration energy spectra for the random solution to values for pure Cr in FCC structure (mimicking the CSRO domain). The barriers are extracted based on the nudged elastic band (NEB) method \cite{henkelman2000climbing}. FIG. \ref{fig:migration}a displays the vacancy migration barriers, which resemble a Gaussian distribution with the following statistics: $E_m^{\mathrm{Fe-V}} = 0.99\pm0.14$ eV, $E_m^{\mathrm{Ni-V}} = 0.97\pm0.18$ eV, and $E_m^{\mathrm{Cr-V}} = 0.84\pm0.14$ eV. The interstitial migration is considered via the rotation of [110] dumbbell \cite{jin2018thermodynamic}. We discard the instances which do not yield a bell curve from NEB calculations due to instability. Those barriers extracted from NEB energy profiles (FIG. \ref{fig:migration}b) indicate that interstitial migration biases toward low-barrier transitions, but still exhibits a significant portion of high barriers. Given the distributions, it can be inferred that Cr is a fast diffuser than Ni and Fe, consistent with previous calculations \cite{bonny2013interatomic} and experimental measurements \cite{million1985diffusion}. In comparison, for pure Cr, we obtained $E_m^{\mathrm{Cr-V}} = 0.84$ eV and $E_m^{\mathrm{Cr-I}} = 0.045$ eV. Therefore, it can be concluded the CSRO domains can cause an apparent change in defect diffusion barriers, particularly for Cr interstitial. Previously, CSRO was demonstrated to raise the activation barriers in dislocation activities \cite{li2019strengthening} due to heightened ruggedness of the PEL. Following the same argument, a higher effective migration barrier is expected due to the existence of CSRO, as also found in MEA/HEA \cite{cao2021does}. Here is a qualitative understanding: although the CSRO domains (Cr-rich) represent a more regular PEL with a high defect diffusivity, such domains can effectively act as traps which limit the long-range migration of defects once they enter such zones, consistent with the trapping/detrapping behavior of dislocation segments during dislocation motion \cite{li2019strengthening,jian2020effects}. Therefore, similar to the coherent precipitates where defects can be temporarily trapped \cite{was2016fundamentals}, a modified defect reaction-diffusion model is necessary to account for the CSRO domains in studying microstructural evolution with continuum methods. 

The radiation behavior is directly elucidated with MD simulations of the radiation damage process \cite{jin2018thermodynamic}. A total of 1,200 random 5\,keV damage cascades are consecutively introduced into the system at 300\,K, with each damage cascade lasting around 30 ps, which is long enough to anneal short-lived defects. This is approximately equivalent to 0.55 dpa (calculated based on the NRT model with a displacement threshold energy 40 eV \cite{astm1996standard,was2016fundamentals}). The high energy collisions are accommodated by splining the EAM potential \cite{bonny2013interatomic} to  Ziegler–Biersack–Littmark potential \cite{ziegler1985stopping} for short-range interactions (splining range 0.35-0.5\,\AA). Note that due to the limited timescale of MD, long-range thermally driven diffusion is ignored. However, the results can provide valuable insights toward defect accumulation and chemical mixing, especially in the scenarios of low-temperature irradiation. The initial systems are corresponding to the three CSRO degrees ($\alpha'$, $\alpha''$, and $\alpha'''$ in FIG. \ref{fig:mc-ordering}). For statistical quality, the results for each CSRO case are averaged over six individual simulations of the damage process. Defects and dislocations are identified with OVITO \cite{stukowski2009visualization}.  

\begin{figure*}[ht!]
  \centering
  \includegraphics[width=0.75\linewidth]{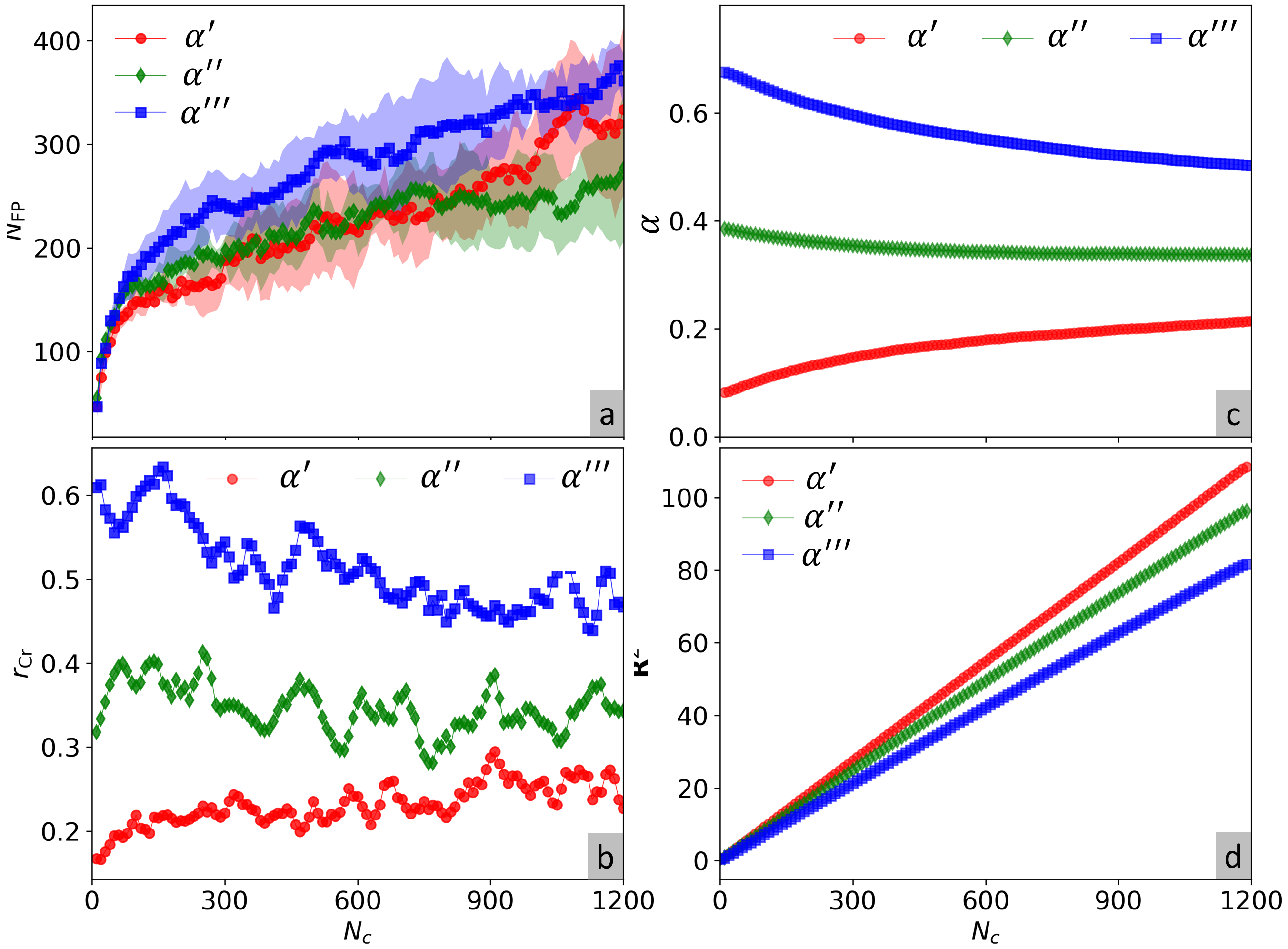}
  \caption{(a) Number of Frenkel pairs ($N_\mathrm{FP}$) with the shaped bands from standard deviation, (b) the fraction of Cr atoms in interstitial clusters ($r_\mathrm{Cr}$), (c) CSRO parameter ($\alpha$), and (d) squared displacements ($\mathbf{R}^2$) versus number of cascades ($N_c$), corresponding to CSRO $\alpha'$, $\alpha''$, and $\alpha'''$, respectively.}
  \label{fig:defects_num}
\end{figure*}
FIG. \ref{fig:defects_num}a records the instantaneous number of Frenkel pairs ($N_{\mathrm{FP}}$) versus the number of damage cascades ($N_c$) or dose for the three CSROs, showing a fast increase in $N_{\mathrm{FP}}$ and gradually approaching saturation. In the initial stage ($N_c\lesssim50$), i) $N_{\mathrm{FP}}$ increases linearly due to the individual damage cascades dispersed in the simulation cell; and ii) the three curves closely overlap, suggesting that the number of defects generated in a single cascade is statistically similar regardless of the CSRO degree. As more cascades are consecutively introduced ($N_c\gtrsim50$), the debris from previous cascades starts to affect defect generation and recombination. The discernible difference in the absolute values of $N_{\mathrm{FP}}$, suggests that defect recombination is less efficient in high CSRO scenarios than random solution. Note that the current simulation cells are defect-sink-free. It means that, without the presence of extended defects such as grain boundaries and dislocations, systems with CSRO do not necessarily show better radiation resistance than random solution in terms of defect concentration. However, realistic systems contain numerous defect sinks to preferentially absorb fast-diffusing interstitials, the temporary interstitial trapping effect enabled by CSRO domains can enhance recombination, hence good radiation tolerance should be expected with CSRO \cite{zhang2017local}.
\begin{figure*}[ht!]
  \centering
  \includegraphics[width=0.65\linewidth]{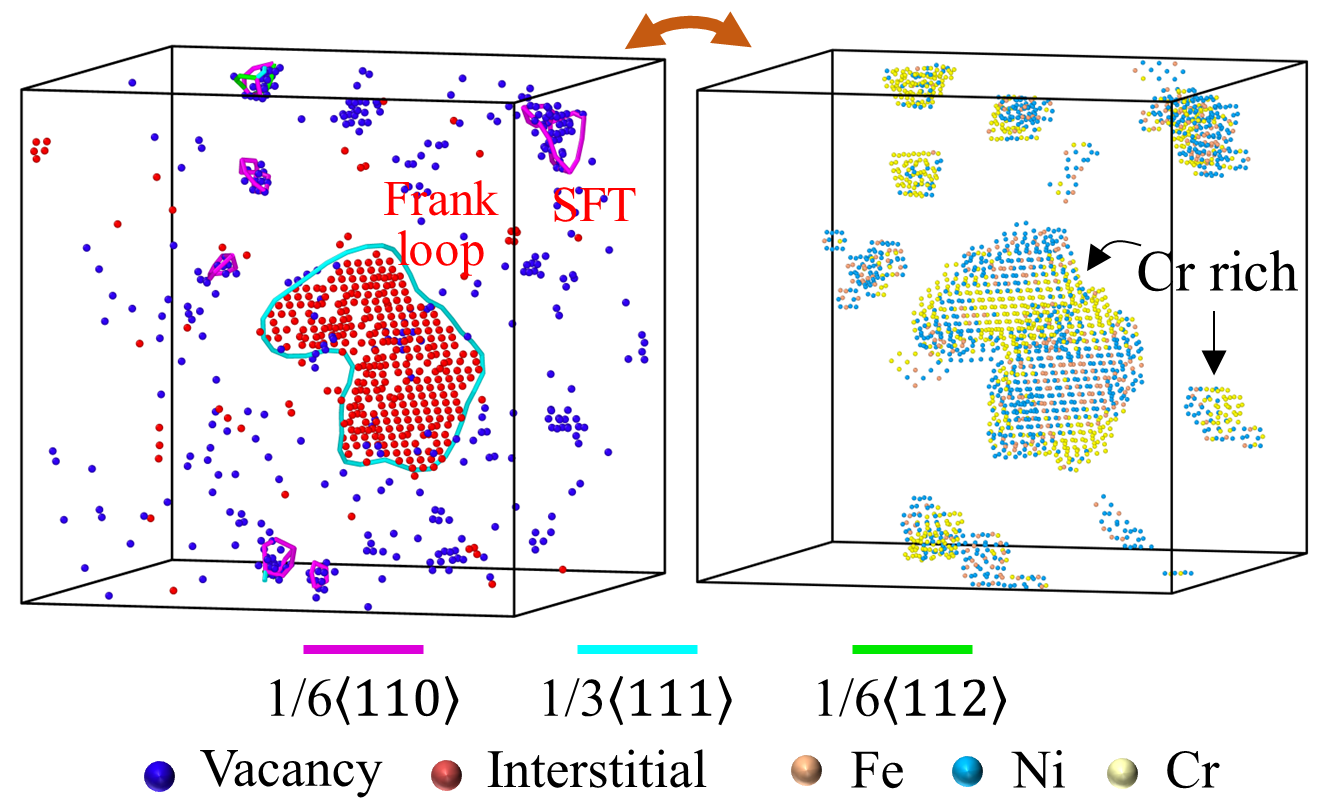}
  \caption{Large interstitial clusters at $N_c=1200$, for the case of CSRO $\alpha'''$. Perfect atoms are removed for visualization purposes.}
  \label{fig:defect_com}
\end{figure*}

\begin{figure*}[ht!]
  \centering
  \includegraphics[width=0.75\linewidth]{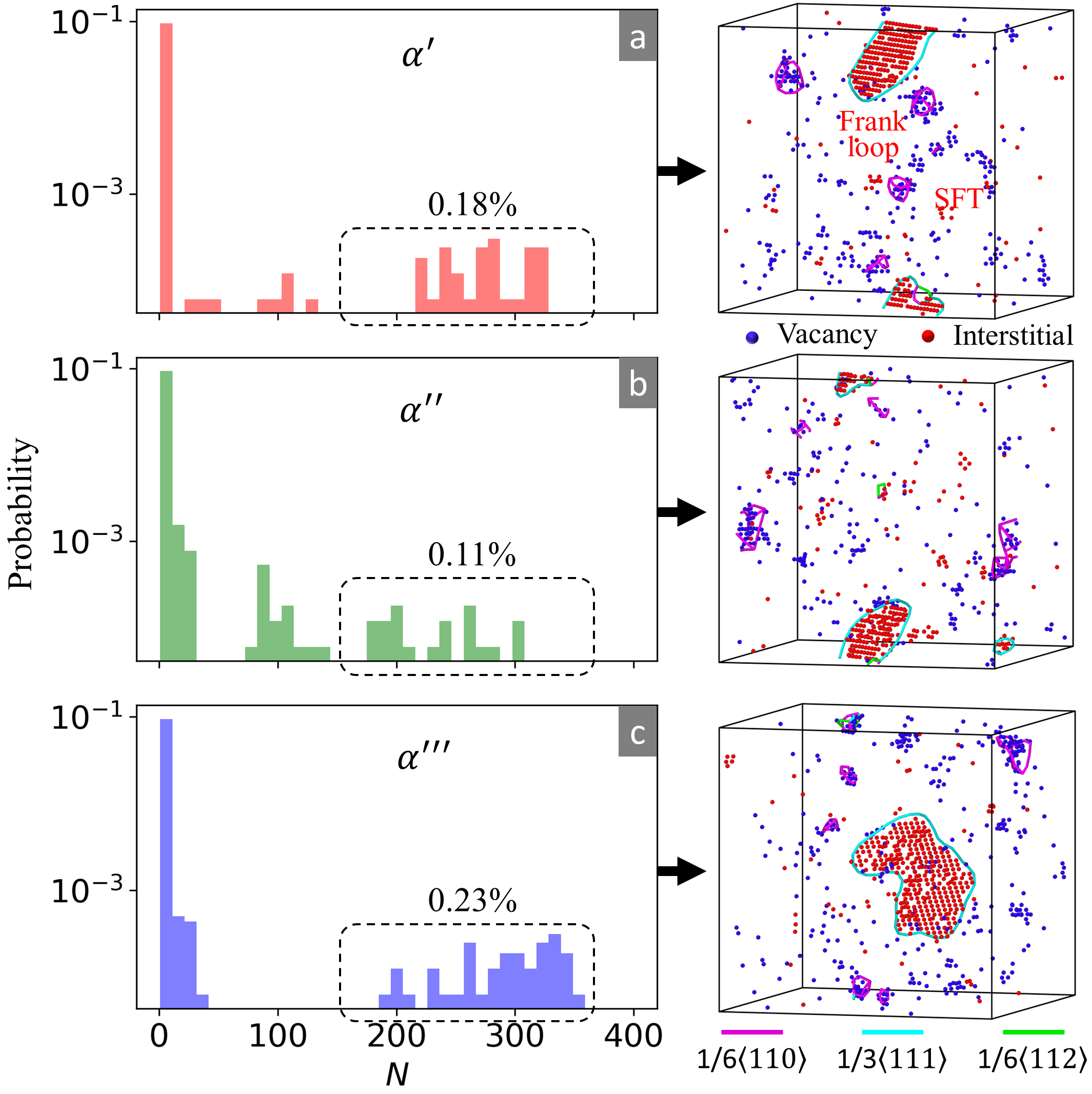}
  \caption{(a-c) Cluster size distribution at $N_c=1200$, corresponding to CSRO $\alpha'$, $\alpha''$, and $\alpha'''$, respectively. The right panel depicts the typical defect distribution and dislocations for each case. Perfect atoms are removed for visualization purposes.}
  \label{fig:defect_dist}
\end{figure*}

Although it is impossible to properly define the nature of vacancies given the dynamic occupancy of constituent elements, the interstitial (cluster) composition can be exactly identified. FIG. \ref{fig:defects_num}b plots the fraction of Cr atoms in interstitial clusters (size$>1$) (denoted as $r_\mathrm{Cr}$) as a function of dose. By comparison, Cr atoms become more involved in interstitial clusters as the CSRO increases. In the case of $\alpha'''$, $r_\mathrm{Cr}$ strongly deviates from the composition ratio (0.3), as defect clusters prefer to locate in/near Cr-rich domains (FIG. \ref{fig:defect_com}). In this case of $\alpha'''$, $\alpha$ is decreasing with radiation (see the next paragraph), $r_\mathrm{Cr}$ exhibits a decreasing trend, suggesting that the $r_\mathrm{Cr}$ is positively correlated with CSRO degree. To gain more insights into the properties of interstitial clusters, we calculate the size distribution at $N_c\approx1200$, as shown in FIG. \ref{fig:defect_dist}. Representative snapshots of defect distribution and dislocation loops are provided on the right panel (see SM for the complete radiation process). The general defect pattern is found to be consistent across all the cases. Vacancies can cluster into stacking fault tetrahedra (SFTs) while interstitials can form both perfect and faulted/Frank dislocation loops. In comparison, the high CSRO state leads to large interstitial clusters in the form of Frank loops (($\mathbf{b}=a/3\langle111\rangle$) in/near Cr-rich domains (FIG. \ref{fig:defect_com}). Although SFE changes from negative to positive (i.e., thermodynamically more difficult to form stacking faults) from $\alpha'$ to $\alpha'''$, these results indicate that growth of Frank loops results from kinetics-driven interstitial accumulation in/near Cr-rich domains. Moreover, these loop sizes appear limited by the nanoscale CSRO features, but should still be observable in experiments \cite{wang2019defect}. For the perfect loops, they exhibit small sizes (from interstitial clustering rather than faulted loop unfaulting) with a parallelpiped morphology and can be highly mobile, which has also been seen in other FCC metals \cite{jin2021dissociated}. Two consequences are identified: i) they can sweep away the vacancies along its path; ii) they can react with other interstitial loops (see SM). In realistic systems, these perfect loops can be absorbed at grain boundaries, contributing to interstitial deficiency in the bulk. Since CSRO can trap dislocation (loop) motion \cite{li2019strengthening}, such retardation is beneficial for defect recombination.     

To reveal how radiation affects the CSRO, we calculated the CSRO as a function of dose (FIG. \ref{fig:defects_num}c). For random solution, the accumulated local atom rearrangement featuring short-range diffusion during the primary radiation stage leads to a gradual increase in CSRO. In the high CSRO scenario, the Cr-rich domains are disrupted by ballistic collisions and thermal spike atom mixing, causing a decreasing trend in CSRO. For the intermediate CSRO degree, radiation changes slightly the order parameter. An asymptotic/steady-state value around 0.34 may be reached if the dose level is significantly increased beyond $N_c=1200$. The existence of steady-state CSRO has also been reported in irradiated Fe-Cr alloy \cite{gomez2014decoupling}. This steady-state value is expected to depend on the irradiation conditions such as irradiation particles, flux, and temperature, but not on the initial structure. Hence, under irradiation, two competing effects are expected. i) Athermal radiation can either increase or decrease CSRO (this work). ii) Radiation-enhanced thermal diffusion can increase CSRO, which has been experimentally identified in various alloys such as Fe-Cr \cite{gomez2014decoupling}, CrCoNi \cite{zhang2017local}, and NiCoFeCrMn \cite{su2022radiation} where CSRO is enhanced by increasing the irradiation dose and temperature. Combining these two, it is projected that the initial CSRO level, dose rate, and temperature determine the evolving trend of CSRO. 

Related to the CSRO change, we examined radiation-induced chemical mixing. Here, we quantify it via the squared displacements $\mathbf{R^2}$ \cite{jin2020achieving} of all atoms in the system with the initial structure as the reference (including atom exchanges within the same type of atoms). It is found that the efficiency of mixing is affected by the CSRO level. FIG. \ref{fig:defects_num}d displays strong linearity of $\mathbf{R^2}$ versus dose, where random solution exhibits the highest mixing rate. As mixing mainly occurs during thermal spikes (local melting), the rate of mixing (or the slope of these curves) should physically correlate with the cohesive energy and atom diffusivity in the liquid (molten) phase. The stronger Cr-Cr binding and supposedly lower diffusion in the molten zone (analogy to the crystalline phase, but remains to be quantified for the amorphous phase) can explain the slower mixing rate for $\alpha'''$ case. 

In summary, we have identified that defect properties and radiation behavior have a strong dependence on the local CSRO in Fe-Ni-Cr alloys. As CSRO is more difficult to observe experimentally than any phase formation \cite{chen2021direct}, the hidden effects of CSRO on defect reaction and diffusion kinetics were mostly neglected in the discussions of radiation-driven microstructural evolution. It is hence of significance to take these nanoscale features into future consideration to explain radiation effects in general CSRO-forming alloys under irradiation environments.

\section*{acknowledgements}
The authors wish to acknowledge the support of the Pennsylvania State University.
\bibliography{science}
\end{document}